\documentclass[12pt]{iopart}

\usepackage{graphicx}
\usepackage{subcaption}

\usepackage[separate-uncertainty=true]{siunitx}

\begin{document}

\title[Liquid concentration profiles of OH and H$_2$O$_2$]{Concentration profiles of OH and H$_2$O$_2$ in plasma-treated water: influence of power, gas mixture and treatment distance}

\author{Anna Lena Schöne$^{1}$, Steffen Schüttler$^{2}$, Talisha Mroß$^{2}$, Niklas Eichstaedt$^{2}$, Judith Golda$^{2}$, Lars Schücke$^{1}$, and Andrew R. Gibson$^{1, 3}$}

\address{$^1$Chair of Applied Electrodynamics and Plasma Technology, Ruhr University Bochum, Bochum, Germany}
\address{$^2$Plasma Interface Physics, Ruhr University Bochum, Bochum, Germany}
\address{$^3$York Plasma Institute, School of Physics, Engineering and Technology, University of York, York, UK}

\ead{schoene@aept.rub.de}
\vspace{10pt}

\begin{abstract}
Plasma liquid interactions are important for a range of applications. For these, H$_2$O$_2$ and OH represent two key reactive species, whose concentrations in liquids need to be controlled for effective application outcomes. Here, a combination of gas and liquid simulations is used to study the concentration profiles of H$_2$O$_2$ and OH in water treated by a radio-frequency-driven plasma jet, with a glass capillary between the electrodes, operated in He with admixtures of water vapour. Simulations are compared with measured H$_2$O$_2$ concentrations and found to be in good qualitative agreement as plasma power and water admixture are varied. Simulation results show that the concentration profiles of H$_2$O$_2$ in the liquid are mainly determined by transport, while those of OH are limited by reactions with H$_2$O$_2$, which consumes OH. For a given plasma operating condition, the concentration and penetration depth of H$_2$O$_2$ increase with plasma treatment time, while those of OH tend to decrease because of the increasing H$_2$O$_2$ concentration. Plasma power, water vapour admixture, and the distance between the jet and the liquid surface all allow for the concentrations of H$_2$O$_2$ and OH to be controlled. The OH delivered from the gas phase to the liquid, and its concentration within the liquid are strongly dependent on the reaction pathways occurring in the effluent region, such that the trends in OH density at the end of the plasma region differ from those in the liquid. While the concentration of OH in the liquid is always much lower than that of H$_2$O$_2$, the ratio of the two species can be controlled over orders of magnitude by varying water admixture and power. The highest selectivity to OH is at low water admixtures, low powers and short treatment times, while the highest selectivity to H$_2$O$_2$ is at high water admixtures, high powers and long treatment times.

\end{abstract}

%
%
%
%
%
\clearpage

\section{Introduction}
Low temperature atmospheric pressure plasmas are well known for their ability to produce reactive oxygen and nitrogen species (RONS) \cite{khlyustova2019important, gorbanev2018analysis, kondeti2018long, lietz2016air, lu_reactive_2016}. They are used and researched for a wide range of applications, such as plasma medicine \cite{keidar2015plasma}, disinfection and sterilisation \cite{laroussi2003nonthermal, gallagher2007rapid, marsit2017terminal}, the production of chemical compounds \cite{gorbanev2020nitrogen} and, recently, plasma-driven biocatalysis \cite{yayci2020microscale, yayci2020plasma, wapshott-stehli_plasma-driven_2022}. All these applications have the challenge that the reactive species produced in the plasma need to be delivered towards a target in a controlled way. In addition, since different reactive species are required for different applications, selective delivery of the desired reactive species is also important. Often times the target is contained within, or consists of, a liquid. In these cases, the reactive species produced in the gas phase need to be delivered into a liquid environment creating a system involving complex plasma-gas-liquid interactions \cite{heirman2024critical, park2021stabilization, brubaker2018liquid, heirman_reactivity_2019}. Even though plasma treatments of liquids are extensively studied, the interactions between the gas phase and liquid phase, and the species evolution within the liquid, are still topics of active research \cite{ikuse2022roles, heirman2024critical, park2021stabilization, semenov_modelling_2019}. In this context, understanding of how to control and optimise them for applications is still developing.

Plasma jets \cite{winter2015atmospheric, reuter2018kinpen, golda_concepts_2016} are suitable sources to provide a variety of reactive species and facilitate an absence of direct interactions between plasma and liquid. In such cases, plasma-generated species are transported from the plasma towards the liquid via an effluent region, using a directed gas flow. Since reactive species, and their delivery from the plasma into the liquid play a major role in the above mentioned applications, the species transport and reactions in the effluent of plasma jets have been extensively investigated by experiment and simulation \cite{herrera_quesada_two-dimensional_2025, labenski_influence_2025, ellerweg_characterization_2010, harris_spatial_2023, myers_atomic_2021, jiang_experimental_2022}. From these studies, it is well established that the densities of certain reactive species, such as OH and O, decay rapidly in the plasma effluent region through chemical reactions \cite{herrera_quesada_two-dimensional_2025, labenski_influence_2025, ellerweg_characterization_2010, myers_atomic_2021, jiang_experimental_2022}. The specific decay rates of these densities have been shown to be sensitive to the composition of the effluent region, as this affects the rate of consumption of each species \cite{herrera_quesada_two-dimensional_2025, labenski_influence_2025, schuttler_production_2024}. On the other hand, less reactive species such as H$_2$O$_2$ and O$_3$ are much more stable and can be transported over longer distances in the effluent \cite{harris_spatial_2023,ellerweg_characterization_2010, jiang_experimental_2022}. The densities of reactive species delivered from the plasma to the effluent, and through the effluent region, are also known to be strongly dependent on the plasma operating conditions, with gas composition and plasma power playing key roles \cite{schroter_chemical_2018, schroter_formation_2020, gianella_ho2_2018, hahn_absolute_2023, gorbanev_combining_2018, jiang_experimental_2022}.

Furthermore, a variety of experimental and modelling studies have focused on understanding species transport and reactions when they enter the liquid environment both in plasma jets and other discharge arrangements \cite{lietz_air_2016, heirman_reactivity_2019, semenov_modelling_2019, sgonina_reactions_2021, poggemann_transportation_2025, verlackt_transport_2018, lindsay_momentum_2015, oinuma_controlled_2020, norberg_atmospheric_2014}. Similarly to the behaviour of species in the effluent region, these studies demonstrate that some species can be transported over comparatively long distances in liquids, while others react quickly to form stable products \cite{heirman_reactivity_2019, oinuma_controlled_2020, norberg_atmospheric_2014, ikuse2022roles}. The rate of transport in liquids treated by jets with an active gas flow incident on the liquid has been demonstrated to be strongly influenced by gas-flow induced convection, as purely diffusive transport is much slower in liquids than in gases \cite{heirman_reactivity_2019, verlackt_transport_2018, lindsay_momentum_2015, kamidollayev_modeling_2023}.

In this work, we build upon previous studies to better understand how the concentrations and penetration depths of OH and H$_2$O$_2$ in liquids treated by plasma jets are dependent upon the plasma operating conditions. OH and H$_2$O$_2$ are studied here largely because of their relevance for plasma-driven biocatalysis. Specifically, plasma-produced H$_2$O$_2$ can be used to drive the enzymatic conversion of substrate molecules into products \cite{yayci2020plasma}. However, OH has the potential to inactivate the enzymes used in the process. In this context, an effective plasma-driven biocatalysis process would aim to optimise the concentration profile of H$_2$O$_2$ within the liquid, while avoiding delivery of significant quantities of OH. While plasma-driven biocatalysis is a key motivator of this work, the results are also expected to be applicable to a wide range of other applications where plasma jets may be used to deliver OH and H$_2$O$_2$ to liquids, such as in plasma medicine, and plasma-driven chemical conversion.

To study the concentration profiles of these species in plasma treated water, simulations of reactive species production and delivery across the effluent are carried out using a 0-D plug-flow model, which is used to derive the fluxes of OH and H$_2$O$_2$ into liquids. These fluxes are used as inputs to a 1-D reaction diffusion model of OH and H$_2$O$_2$ reactions and transport in the liquid, which allows for their concentration profiles as a function of depth within the liquid to be simulated. This approach is used to understand how the concentration profiles of each species can be controlled by tailoring the plasma conditions, including plasma power, water vapour admixture in the gas phase, as well as the distance between the jet and the liquid surface. The model is also compared with experimentally measured H$_2$O$_2$ concentrations in liquids. Section \ref{sec:setup} describes the experimental setup and the models used, while the results and conclusions are presented in sections \ref{sec:results} and \ref{sec:conclusions}, respectively.

\section{Setup} \label{sec:setup}
An atmospheric pressure plasma jet (APPJ) is used in this work for liquid treatments. It is driven by a radio-frequency (RF) voltage source, and by applying a gas flow through the APPJ, the species produced in the plasma are transported into a liquid via its effluent, where gas-liquid interactions determine the amount of dissolved species. 

This section first describes the simulation of the plasma liquid system and, second, the experimental setup for validating the simulation results.

\subsection{Computational model} \label{sec:model}
\begin{figure}[htb]
    \centering
    \includegraphics[width=0.4\textwidth]{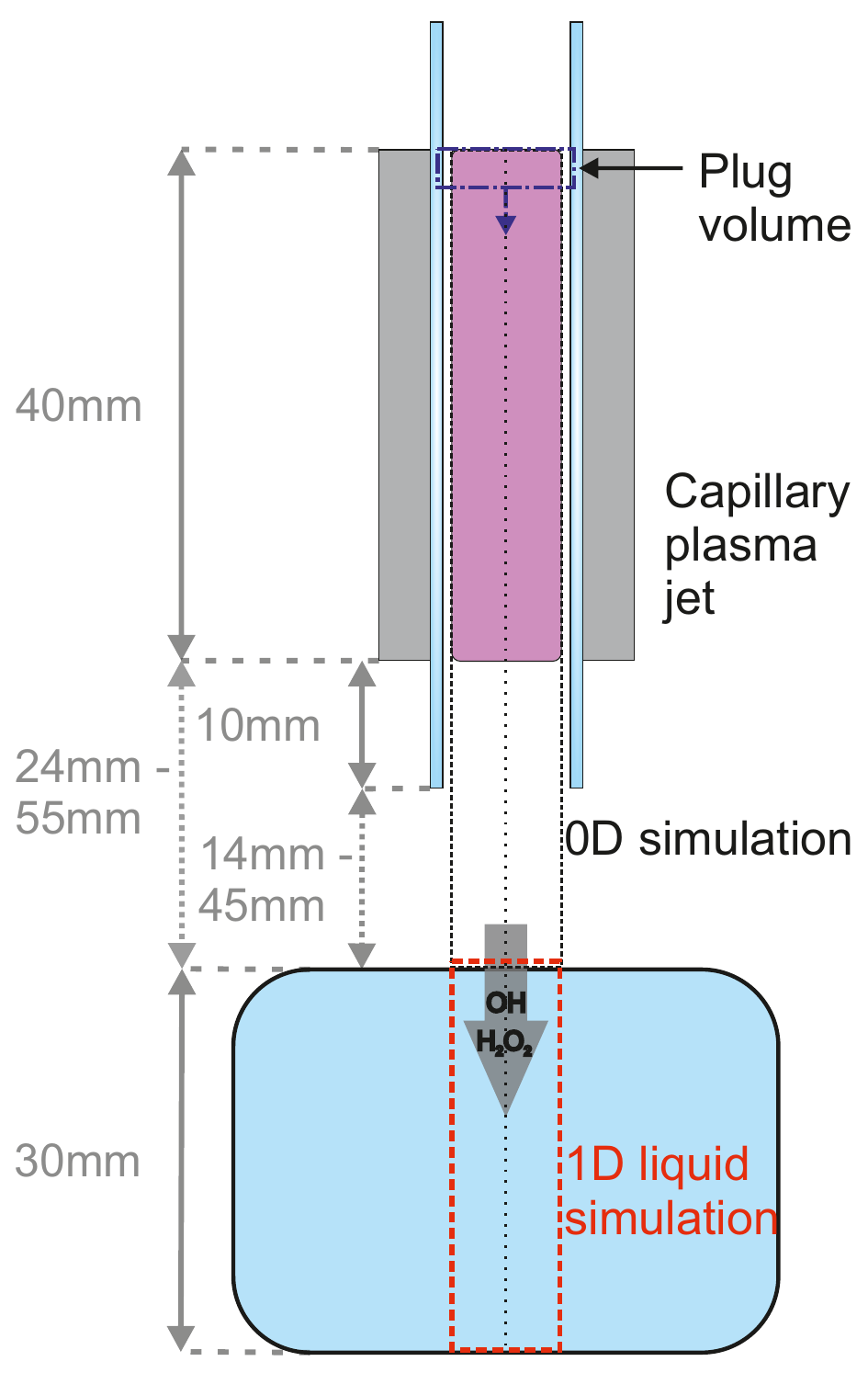}
    \caption{Schematic drawing of the simulated and experimental configuration. 0-D plasma chemical kinetic simulations, assuming a plug-flow model, are used for the gas phase region (plasma and effluent). 1-D simulations based on the reaction-diffusion equation are used for the liquid phase region.}
    \label{fig:setup}
\end{figure}

The simulation setup is divided into gas and liquid phase simulations as depicted in Fig.~\ref{fig:setup}. This approach is previously used and described in \cite{poggemann_transportation_2025}, so only a brief summary is given here.

The gas phase simulation is performed using GlobalKin \cite{lietz_air_2016}, a zero-dimensional plasma chemical kinetics code with an integrated plug-flow model that calculates spatially dependent values along the gas flow direction using the gas flow velocity as a conversion factor. Therefore, it allows the spatial variation of species densities along the plasma channel and into the effluent region up to the liquid surface to be simulated. The base case dimensions of the simulated gas phase setup are 40\,mm long electrodes, followed by an additional 10\,mm of capillary-guided effluent, and 14\,mm unguided effluent until the liquid surface. In the first 40\,mm, power is deposited between the electrodes, forming the plasma region. The following 24\,mm are simulated without power input, so that no active plasma is present. The chemistry of the gas phase is simulated using a helium-water vapour reaction scheme \cite{brisset_chemical_2021}, which was developed and validated for jet configurations without dielectric materials between the electrodes \cite{brisset_chemical_2021,schroter_chemical_2018,schroter_formation_2020,schroter_numerical_2017}, and has also been used to study the capillary jet used in this work \cite{schuttler_production_2024,poggemann_transportation_2025}.

In order to account for the chemistry in an unguided effluent within the plug-flow model, assumptions about the gas velocity in this region are necessary. Here, it is assumed that the gas flow velocity in the unguided effluent reduces linearly from the end of the capillary to the liquid surface. Hence, the simulation length is adjusted to a longer distance to allow for an equivalent residence time in this region. This results in the assumption of a flow velocity of $v_{flow}\,=\,$0 at the liquid surface $z\,=\,$0.

Overall, the gas phase simulations result in spatially dependent species behaviour in the plasma channel and the effluent up to the liquid surface. The species densities calculated in the gas phase directly above the liquid surface are used as an input for the liquid simulations.

The behaviour of H$_2$O$_2$ and OH in the liquid is calculated by solving the reaction-diffusion equation for one dimension in space, using pdepe in MATLAB \cite{pdepdeMATLAB}, as described in \cite{poggemann_transportation_2025}. The initial liquid consists of nearly pure water with concentrations of 1.661$\cdot10^{-17}\,$M (1$\cdot10^{10}\,$m$^{-3}$) of H$_2$O$_2$ and OH. At the bottom of the liquid, a constant flux boundary condition is applied. At the top of the liquid surface, H$_2$O$_2$ and OH fluxes from the gas phase into the liquid phase are calculated based on their densities from the gas phase simulations. The expressions used to determine the flux of species into the liquid from the gas phase are described in detail in \cite{poggemann_transportation_2025}, and are largely derived from the work of Semenov \emph{et~al.} \cite{semenov_modelling_2019}. These fluxes are dependent on a range of factors, including the thicknesses of the diffusion boundary layers at the gas and liquid sides of the interface, the gas and liquid phase diffusion coefficients of the species, and their Henry's solubility constants.

A simplified reaction scheme is used to describe H$_2$O$_2$ and OH reactions in the liquid. This includes two reactions: 
\begin{equation}
    \mathrm{OH\,+\,OH\,\rightarrow{}\,H_2O_2}
\end{equation}
\begin{equation}
    \mathrm{OH\,+\,H_2O_2\,\rightarrow{}\,HO_2\,+\,H_2O}
\end{equation}

For solution of the reaction-diffusion equation in the liquid, diffusion coefficients are required for both species. For the jet system studied in this work, it was demonstrated in \cite{poggemann_transportation_2025} that the transport of H$_2$O$_2$ in the liquid is not well described when only diffusive transport is accounted for. This is due to gas-flow induced convection, which drives faster transport than diffusion alone. The relative importance of convective transport scales with the gas flow velocity. At a gas flow velocity of 0.25\,slm it was found that the rate of H$_2$O$_2$ transport could be accurately simulated by using an experimentally derived effective diffusion coefficient in the reaction-diffusion equation \cite{poggemann_transportation_2025}, instead of the molecular diffusion coefficient. This effective diffusion coefficient represents a combination of diffusive and convective transport. In this work, we focus on simulating cases with a gas flow rate of 0.25\,slm and use the experimentally derived effective diffusion coefficient from \cite{poggemann_transportation_2025} in the reaction-diffusion equation for H$_2$O$_2$. For OH, the molecular diffusion coefficient derived from molecular dynamics simulations in \cite{poggemann_transportation_2025} is used. This is a more reasonable approximation for OH because of its much higher reaction rate in liquids, which means that its penetration into liquids is more limited by reactions rather than its rate of transport. Further discussion on the assumptions around species transport used in the model are given in \cite{poggemann_transportation_2025}.

The liquid simulations result in species concentration profiles of the considered species as functions of the depth into the liquid and the treatment time. Simulations are performed for various operating conditions to investigate the influence of the deposited plasma power, the water admixture in the feed gas, and the effluent region's length on the concentrations of H$_2$O$_2$ and OH in the liquid environment.

\subsection{Experimental setup}
The APPJ used here is based on the design of the well-investigated COST reference plasma jet, extended by the use of a borosilicate glass capillary (CM Scientific) acting as a dielectric. The length of the stainless steel electrodes was 40\,mm, and the inner square cross-section of the square capillary was 1\,mm$^2$. The distance between the electrodes was 1.4\,mm due to the thickness of the capillary walls which was 0.2\,mm. The power was provided by an RF power generator (Coaxial Power Systems RFG 150) operating at a frequency of 13.56\,MHz. It was connected to one electrode via a matching network (Coaxial Power Systems MMN 150), while the second electrode was grounded. Power measurements were performed as described in previous publications \cite{golda_dissipated_2019, schuttler_production_2024}. The signals of voltage and current were measured by a 10\,GS/s oscilloscope (Teledyne LeCroy HDO6104A) to obtain a sufficient time resolution for phase detection between the signals. Mass flow controllers (Analyt MTC) were used to control the gas flow, which was set to 0.25\,slm. The plasma jet was operated in helium (purity 5.0), and the use of an ice-cooled bubbler system filled with distilled water enabled the addition of water vapour to the helium gas flow. The temperature in the bubbler vessel was controlled to \SI{1.40\pm0.22}{\celsius}, resulting in a maximum water addition of \SI{6400\pm240}{ppm}. Liquid treatments were performed by treating 3$\,$mL of liquid in a UV cuvette (Sarstedt polystyrene). The default distances between the plasma and the end of the capillary to the liquid surface were 24\,mm and 14\,mm, respectively. Thus, the effluent was guided for 10\,mm through the capillary after the plasma region, protecting it from mixing with the atmosphere. For 14\,mm, the effluent passed through the atmosphere until it reached the liquid surface. The treatment time was 900\,s during which evaporation was measured to be less than 3\,\%. This evaporation is neglected in the analysis of the results.

The H$_2$O$_2$ concentration was measured by spectrophotometry using the addition of ammonium metavanadate to the treated liquid as described in a previous work \cite{schuttler_validation_2023}. Ammonium metavanadate reacts with H$_2$O$_2$, producing a red-orange peroxovanadium cation solution with an absorption peak at 450$\,$nm. The dissolution of H$_2$O$_2$ in the ammonium metavanadate solution shows the same behaviour as a buffered solution, as shown in a previous work \cite{schuttler_validation_2023}. Thus, we assume that the results obtained with the ammonium metavandate solution show similar behaviour to plasma-treated water or buffered solutions. Light from a laser-stabilised broadband light source (Energetiq EQ-99 LDLS) was guided through the cuvette, and the absorption spectra were measured by a spectrometer (Avantes Avaspec-ULS 2049x64 TEC-EVO). Background and reference spectra were recorded before each treatment. The measurements took place at a depth of 3$\,$mm to 6$\,$mm beneath the liquid surface and were performed after plasma treatment and after stirring the treated solution to ensure good mixing. Calibration of the spectrophotometric setup was performed by known H$_2$O$_2$ solutions with concentrations up to 1$\,$mM and linear behaviour of the absorption with H$_2$O$_2$ concentration was observed. The uncertainty of the H$_2$O$_2$ measurements in the treated liquid was estimated from three consecutive measurements at standard parameters of 6\,W and 6400\,ppm, showing a standard deviation of about 14\,\%. This value was assumed as maximum uncertainty for all measurements.

\section{Results and discussion} \label{sec:results}
\begin{figure}[htb]
    \centering
    \includegraphics[width=1.0\textwidth]{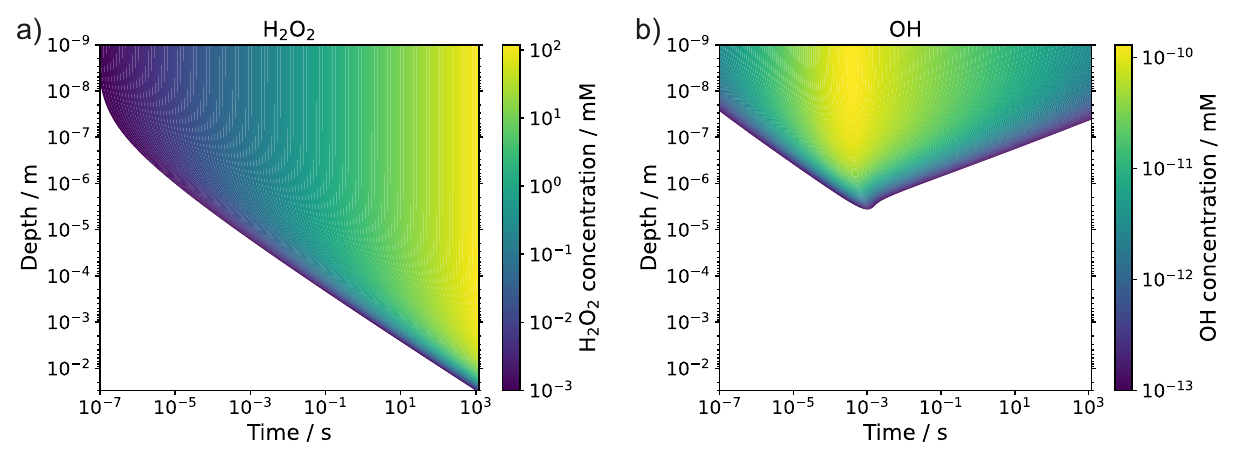}
    \caption{Spatio-temporal profiles of a) H$_2$O$_2$ and b) OH for a plasma power of 6.0\,W, a water admixture of 6400\,ppm, and a distance between the end of the plasma forming region and the liquid surface of 24\,mm.}
    \label{fig:contour}
\end{figure}

The calculation of the species behaviour in the liquid using the reaction-diffusion equation results in spatio-temporal profiles of H$_{2}$O$_{2}$ and OH, as shown in Fig.~\ref{fig:contour} for the base case conditions of this study.

The H$_{2}$O$_{2}$ concentration distribution (Fig.~\ref{fig:contour}~a)) is mainly determined by its transport in the liquid, indicated by a consistent evolution over time. More specifically, the H$_{2}$O$_{2}$ concentration increases gradually in time, while the expansion in depth proceeds homogenously up to a boundary where the concentration drops sharply. The rate of expansion of the H$_{2}$O$_{2}$ concentration into the liquid is determined by the rate of transport, which in this work is defined by the experimentally measured effective diffusion coefficient from \cite{poggemann_transportation_2025}, as described in section \ref{sec:model}. Simulations using this approach with the effective diffusion coefficient were found to be in good agreement with time and space resolved measurements of relative H$_{2}$O$_{2}$ concentrations in liquids in \cite{poggemann_transportation_2025}.

The concentration of OH (Fig.~\ref{fig:contour}~b)) first shows a comparable behaviour to H$_{2}$O$_{2}$ up to the range of milliseconds, albeit with significantly lower concentration values, followed by a decrease of the OH concentration and the penetration depth with time. It is mainly located at the liquid surface, never penetrating with any significant concentration beyond a few $\mu$m from the interface with the gas phase. The consumption of OH during chemical reactions causes this rapid decrease.

In the following, the influence of the plasma operating conditions on the concentrations of H$_{2}$O$_{2}$ and OH in the liquid is investigated. Specifically, the water admixture to the feed gas, the deposited plasma power, and the distance between the capillary jet and the liquid surface are varied.

\subsection{Influence of plasma parameter variation on the H$_2$O$_2$ behaviour}
\begin{figure}[htb]
    \centering
    \includegraphics[width=1.0\textwidth]{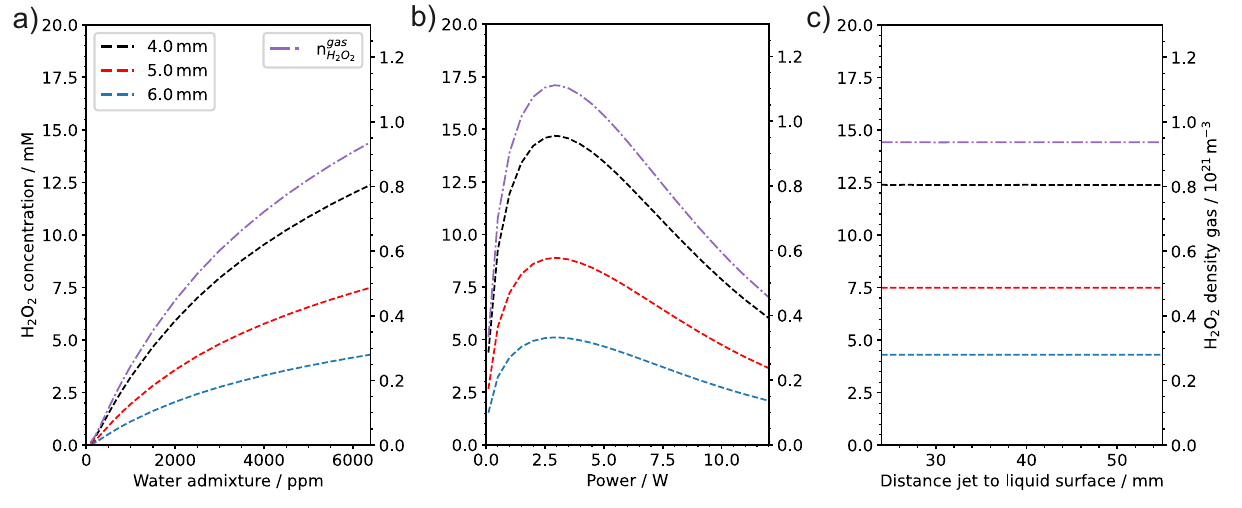}
    \caption{Simulated H$_2$O$_2$ behaviour in the liquid over a) water admixture, b) deposited plasma power, and c) distance between jet and liquid surface (for 4.0\,mm, 5.0\,mm, and 6.0\,mm below the liquid surface), compared to the simulated gas phase behaviour. Base conditions are at 6.0\,W deposited plasma power, 6400\,ppm water admixture, a distance between the end of the plasma forming region and the liquid surface of 24\,mm and 300\,s treatment time. Legend is true for all.}
    \label{fig:H2O2_compGas}
\end{figure}

The influence of varying a) the water admixture to the feed gas, b) the deposited plasma power, and c) the distance between the capillary jet and the liquid surface on H$_2$O$_2$ is shown in Fig.~\ref{fig:H2O2_compGas}. Here, the H$_2$O$_2$ concentrations in the liquid 4.0\,mm, 5.0\,mm and 6.0\,mm below the liquid surface are compared to the H$_2$O$_2$ densities in the gas phase ($n_{H2O2}^{gas}$) directly above the liquid, that serve as an input for the liquid simulation.

In the case of the water admixture variation (Fig.~\ref{fig:H2O2_compGas}~a)), the influence of the water content of the feed gas is investigated for nearly dry helium (100\,ppm water admixture) up to saturation of water vapour in the helium feed gas at 6400\,ppm. Here, the H$_{2}$O$_{2}$ density in the gas phase increases with water admixture. First, nearly linearly up to 1000\,ppm, followed by a slower increase resulting in a H$_{2}$O$_{2}$ density of 0.94$\cdot10^{21}\,$m$^{-3}$ at 6400\,ppm. Comparable behaviour is found for the H$_{2}$O$_{2}$ concentration in the liquid, while the admixture trend is not strongly dependent on the depth in the liquid. The H$_{2}$O$_{2}$ concentration decreases with increasing depth in the liquid from a maximum at 6400\,ppm water admixture of around 12.4\,mM at 4.0\,mm below the liquid surface to around 4.7\,mM at 6.0\,mm below the liquid surface. The H$_{2}$O$_{2}$ concentration in the liquid and the H$_{2}$O$_{2}$ density in the gas phase follow the same trend. Therefore, the gas phase behaviour of H$_{2}$O$_{2}$ allows for a good estimation of the H$_{2}$O$_{2}$ behaviour in the liquid under water admixture variation.

For the variation of the deposited plasma power (Fig.~\ref{fig:H2O2_compGas}~b)), the gas phase density of H$_{2}$O$_{2}$ shows an increase up to 3\,W with a maximum H$_{2}$O$_{2}$ density of around 1.11$\cdot10^{21}\,$m$^{-3}$ followed by a decrease with increasing power. Again, the liquid simulation results behave similarly, with similar trends at different depths with the H$_{2}$O$_{2}$ concentration increasing up to around 3\,W, followed by a decrease with increasing power. The concentration of H$_{2}$O$_{2}$ decreases with increasing depth in the liquid from around 14.6\,mM at 4.0\,mm to 5.1\,mM at 6.0\,mm below the liquid surface. As before, the gas phase behaviour allows for a good estimation of the liquid behaviour of H$_{2}$O$_{2}$ with variation of the deposited plasma power.

The H$_{2}$O$_{2}$ remains unchanged in both the gas phase and in the liquid for a variation of the distance between the capillary jet and the liquid surface. With increasing depth in the liquid, the H$_{2}$O$_{2}$ concentration decreases from around 12.4\,mM at 4.0\,mm to 4.3\,mM at 6.0\,mm, while the H$_{2}$O$_{2}$ density in the gas phase is around 0.93$\cdot10^{21}\,$m$^{-3}$.

Overall, the gas phase behaviour of H$_{2}$O$_{2}$ is found to be strongly correlated to its behaviour in the liquid over the different variations carried out here. The gas phase density, therefore, determines the H$_{2}$O$_{2}$ behaviour in the liquid and underlines the assumption of mainly transport-driven behaviour of H$_{2}$O$_{2}$ within the liquid.

\begin{figure}[htb]
    \centering
    \includegraphics[width=1.0\textwidth]{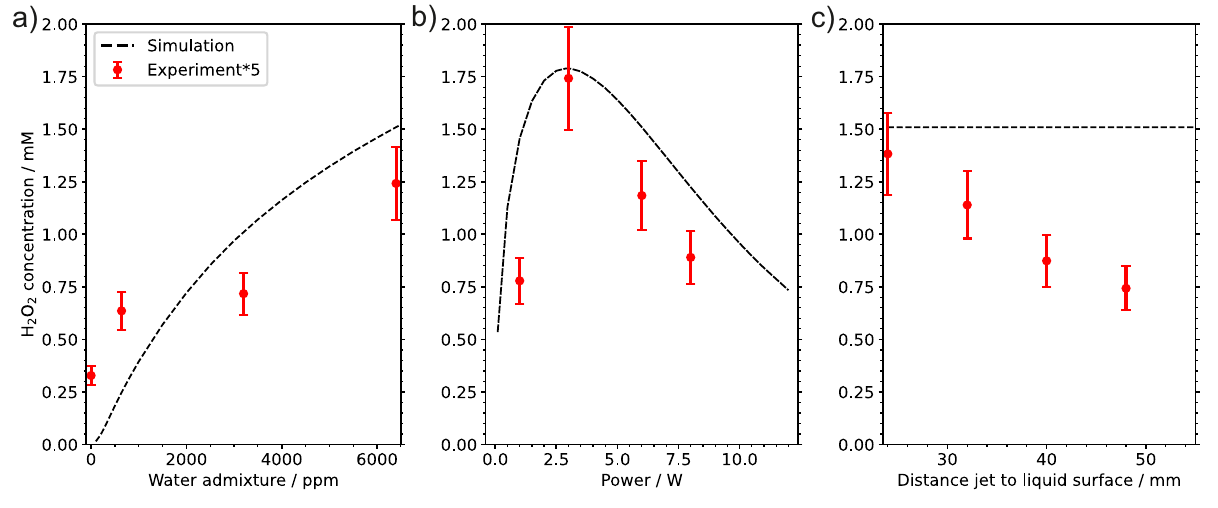}
    \caption{H$_2$O$_2$ behaviour in the liquid over a) water admixture, b) deposited plasma power, and c) distance between jet and liquid surface, integrated over the measured volume compared to the experiment. Base conditions are at 6.0\,W deposited plasma power, 6400\,ppm water admixture, a distance between the active plasma region and the liquid surface of 24\,mm after 900\,s treatment time. Legend is true for all.}
    \label{fig:H2O2_compExp}
\end{figure}

The liquid simulation results for the H$_{2}$O$_{2}$ concentration are compared to experimental measurements for varying water admixture, plasma power and distance between the jet and the liquid surface. Comparisons are carried out after 900\,s treatment time for each case. Since the liquid is stirred before carrying out the measurement of H$_{2}$O$_{2}$ concentration in the experiment, the experimental values represent average concentrations over the whole volume of the cuvette. While the 1-D liquid simulations provide the H$_{2}$O$_{2}$ profile with depth (z-direction) from the liquid surface, they do not provide any information about the H$_{2}$O$_{2}$ concentration profile along the breadth (i.e. in the x- and y-direction) of the cuvette, so this needs to be assumed in order to compare with the experimental measurements. To make the comparison, the simulation results are first integrated with respect to depth, and divided by the total depth of the cuvette to give an average H$_{2}$O$_{2}$ concentration. A Gaussian profile is assumed for the distribution in x- and y-directions. Here, the simulated average H$_{2}$O$_{2}$ concentration is used as the maximum value in the middle of the cuvette decreasing in both directions with a full width half maximum (FWHM) of 3\,mm over a total extent of 10\,mm in both directions. The FWHM is an estimate of the expected width of the H$_{2}$O$_{2}$ profile in the gas phase above the liquid surface. This approximation is based on the fact that H$_{2}$O$_{2}$ is long lived in the gas phase, so it should spread with the gas flow as it exits the capillary. Harris \emph{et~al.} have measured the widths of the H$_{2}$O$_{2}$ distribution exiting the COST jet under a higher gas flow rate than used here \cite{harris_spatial_2023}. In that work, the width of the density distribution was found to be in the mm range, increasing with distance from the exit of the jet. Since the gas flows used in this work are lower, the values from Harris are not used directly, but provide a broad basis from the 3\,mm used here. The integrated simulation results compared to the experimental measurements are shown in Fig.~\ref{fig:H2O2_compExp} for a variation of a) water admixture to the feed gas, b) deposited plasma power, and c) distance between the capillary plasma jet and the liquid surface.

The simulated and measured trends in H$_{2}$O$_{2}$ concentration are in good agreement for variation of the plasma power and the water admixture to the feed gas. On the other hand, the simulations do not reproduce the experimentally measured trend in H$_{2}$O$_{2}$ concentration with varying distance between the end of the active plasma region and the surface. Overall, the absolute values of the experimentally measured  H$_{2}$O$_{2}$ concentration are about a factor of five lower compared to the simulation. Therefore, the simulation overestimates the H$_{2}$O$_{2}$ concentration in the liquid. The overestimation is likely to be caused by simplifying assumptions in the simulations. For example, the spreading of the gas flow between the jet exit and the liquid surface is not accounted for in the simulation, while in the experiment it would mean that not all H$_{2}$O$_{2}$ produced in the gas phase will enter the liquid. In addition, the simulated average concentration used to compare with experiment is also sensitive to the FWHM of the Gaussian distribution that is assumed for the H$_{2}$O$_{2}$ concentration. The value of 3\,mm that is chosen here is a coarse assumption based on a lack of data for H$_{2}$O$_{2}$ density distributions in the gas or liquid phases for the specific conditions used here, and represents a significant uncertainty in the comparison. For the water admixture variation, shown in a), the H$_{2}$O$_{2}$ concentration first increases nearly linearly, followed by a slower increase. Compared to the previous results without integration over the volume, the absolute H$_{2}$O$_{2}$ concentration is lower. This is to be expected as the H$_{2}$O$_{2}$ concentration is not expected to be homogeneous throughout the cuvette, so that when the liquid is stirred the average H$_{2}$O$_{2}$ concentration is lower than the maximum H$_{2}$O$_{2}$ concentration would be before stirring. This is reflected in the simulation by the assumed Gaussian distribution towards the sidewalls of the cuvette in the x- and y-directions. The experimental measurements multiplied by a factor of five are slightly above the simulated H$_{2}$O$_{2}$ concentrations at 0\,ppm and 640\,ppm water admixture. In contrast, the measured H$_{2}$O$_{2}$ concentration at a water admixture of 3200\,ppm and 6400\,ppm lies below the simulated values. Overall, the simulated and experimentally measured trends are in good agreement, considering the above-mentioned assumptions.

The H$_{2}$O$_{2}$ concentration under power variation follows the trend described for Fig.~\ref{fig:H2O2_compGas}. The H$_{2}$O$_{2}$ concentration increases until it reaches a maximum at 3\,W followed by a decrease with increasing deposited plasma power. The trends of simulated and experimentally measured values are in good agreement even though the absolute values differ by a factor of five.

The integrated H$_{2}$O$_{2}$ concentration over the liquid volume of the cuvette also stays constant with increasing distance between the end of the capillary plasma jet and the liquid surface as described for Fig.~\ref{fig:H2O2_compGas}. In contrast, the experimentally measured H$_{2}$O$_{2}$ concentration decreases with an increasing distance to end of the jet. The difference in the trends is expected to be caused by the widening of the gas flow when it exits the capillary and mixes with the ambient air, which is a constant factor for the other parameter variations since the distance between the end of the plasma forming region is always constantly set at 24\,mm. As the distance between the end of the capillary and the liquid is increased, the widening of the gas flow is also expected to increase, which should lead to increased transport of the H$_{2}$O$_{2}$ to the sides, leading to a lower fraction of the available H$_{2}$O$_{2}$ reaching the liquid in the experiment. Since such dynamics are not included in the model, it is reasonable to expect the trends in the H$_{2}$O$_{2}$ concentration to differ in this case.

\subsection{Influence of plasma parameter variation on the OH behaviour}
In the following, the behaviour of OH in the gas and liquid is investigated under variation of the plasma operating conditions. Because the OH concentration in the liquid varies strongly with depth, both the averaged OH concentration and its penetration depth are used to analyse its behaviour. The penetration depth is defined here as the position where the OH concentration reaches a value of $\frac{1}{e}$ of the value directly below the liquid surface.

\begin{figure}[htb]
    \centering
    \includegraphics[width=1.\textwidth]{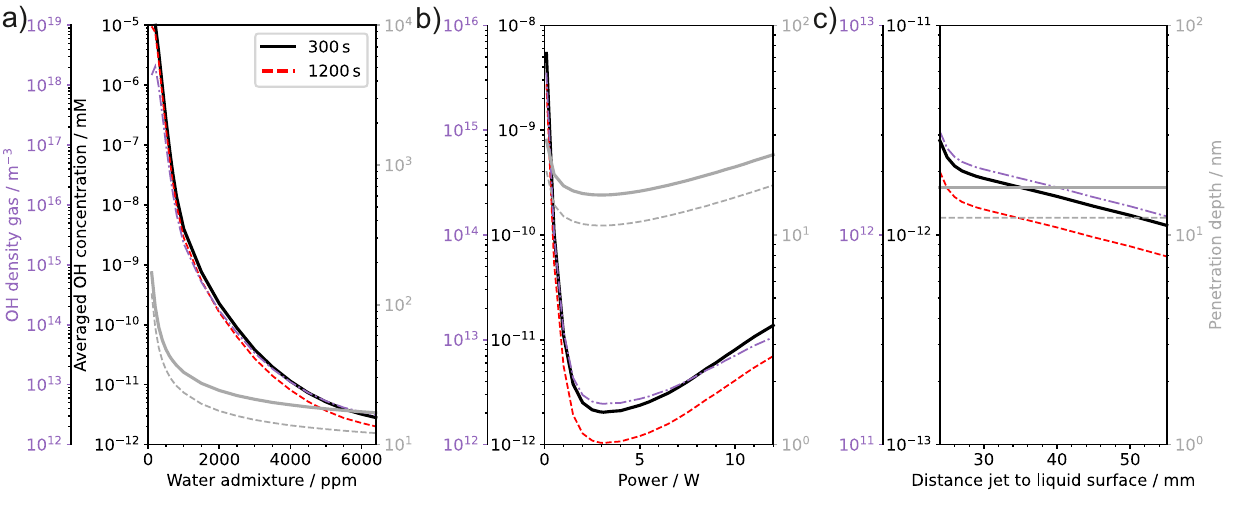}
    \caption{Averaged OH concentration in the liquid, penetration depth of OH in the liquid, and OH density at the gas phase directly above the liquid surface. These are shown as a function of a) water admixture, b) deposited plasma power, and c) distance between jet and liquid surface at 300\,s (solid lines) and 1200\,s (dashed lines). Simulations are carried out at 6.0$\,$W deposited plasma power, and a distance from the end of the active plasma region to the liquid surface of 24$\,$mm.}
    \label{fig:OH}
\end{figure}

The influence of a) the water admixture, b) the deposited plasma power, and c) the distance between the active plasma region and the liquid surface on the gas phase density of OH, the OH concentration in the liquid and the penetration depth of OH in the liquid after 300\,s, and 1200\,s is shown in Fig.~\ref{fig:OH}. The concentration in the liquid is averaged over the distance between the liquid surface and the penetration depth. The OH density in the gas phase is used as input for the liquid simulation and does not depend on the treatment time. Therefore, the OH density in the gas phase is equal for both treatment times.

The results for varying water admixture from nearly dry helium (100\,ppm water admixture) up to saturation at 6400\,ppm are shown in Fig.~\ref{fig:OH}~a). The gas phase density of OH behaves similarly to the OH concentration in the liquid averaged over the penetration depth. The averaged OH concentration and penetration depth both decrease as treatment time is increased from 300\,s to 1200\,s. 
Therefore, the total amount of OH in the liquid also lowers with treatment time. The decrease in penetration depth and averaged OH concentration in the liquid with increasing treatment time are caused by an increase in the H$_2$O$_2$ concentration in the liquid (shown in Fig.~\ref{fig:contour}~a), and discussed in more detail in Sec.~\ref{sec:profiles}). The increase in H$_2$O$_2$ concentration leads to an increase in the consumption of OH in the liquid through reactions with H$_2$O$_2$, which represents the main loss channel of OH in the model. Both the averaged OH concentration over penetration depth in the liquid and the OH density in the gas phase first increase from 100\,ppm to 200\,ppm water admixture reaching maxima of 2$\cdot10^{18}\,$m$^{-3}$ in the gas phase, and 1.1$\cdot10^{-5}\,$mM after 300\,s plasma treatment in the liquid. Above 200\,ppm water admixture, the OH density in the gas phase and the averaged OH concentration over penetration depth in the liquid decrease with water admixture. This decrease becomes less pronounced at higher water admixtures. The penetration depth decays steeply from around 170\,nm after 300\,s treatment time and 120\,nm after 1200\,s treatment time at 100\,ppm water admixture and flattens with increasing water admixture until a penetration depth of around 17\,nm and 12\,nm at a water admixture of 6400\,ppm. The change in penetration depth with increasing water admixture is correlated with the increase in H$_2$O$_2$ concentration in the liquid, which acts to consume OH and limits its penetration depth. The penetration depths for OH found in this work are broadly consistent with those in previous works \cite{gopalakrishnan_solvated_2016, rumbach_penetration_2018}.

The variation of the deposited plasma power shown in Fig.~\ref{fig:OH}~b) also shows similar trends for the OH density in the gas phase and the averaged OH concentration over penetration depth in the liquid. Here, the averaged OH concentration after a treatment time of 300\,s is higher than after 1200\,s. The deviation first increases with deposited plasma power until 3\,W and stays constant afterwards. The OH density in the gas phase, the averaged OH concentration in the liquid, and the penetration depth decrease steeply until 3\,W, followed by a comparably slight and continuous increase. As before, the general trend in averaged OH concentration in the liquid is largely defined by the density in the gas phase at the liquid surface, while the penetration depth and variation of OH concentration with treatment time are related to reactions with H$_2$O$_2$ in the liquid.

In the case of the distance variation between the end of the active plasma region and the liquid surface, the gas phase density of OH and the OH concentration averaged above the penetration depth in the liquid, again, follow similar trends. Both decrease slightly more steeply as the distance between the capillary plasma jet and the liquid surface is varied from 24\,mm to 26\,mm. Above 26\,mm, the OH decreases more slowly, but continuously in the gas phase and the liquid. The averaged OH concentration over penetration depth in the liquid and the penetration depth into the liquid after 300\,s are higher than after 1200\,s of plasma treatment. The distance variation between the active plasma region and the liquid surface does not influence the penetration depth of OH into the liquid. This is due to the constant H$_2$O$_2$ concentration in the liquid phase simulations over this variation. Since the OH concentration averaged over the penetration depth decreases, the total amount of OH in the liquid also decreases.

Overall, all investigated parameter variations influence the behaviour of OH in both the gas and the liquid. The liquid phase concentrations always largely follow the trends of the gas phase densities. The OH densities and concentrations are influenced and vary over around multiple orders of magnitude under the water admixture variation. The deposited plasma power and the distance between the active plasma region and the liquid surface also influence the OH densities and concentrations significantly, where their variations are less, but still significant. In order to better understand the reasons for these trends, the evolution of the chemistry in the gas phase should be considered.

\begin{figure}[p]
    \centering
    \includegraphics[width=1.\textwidth]{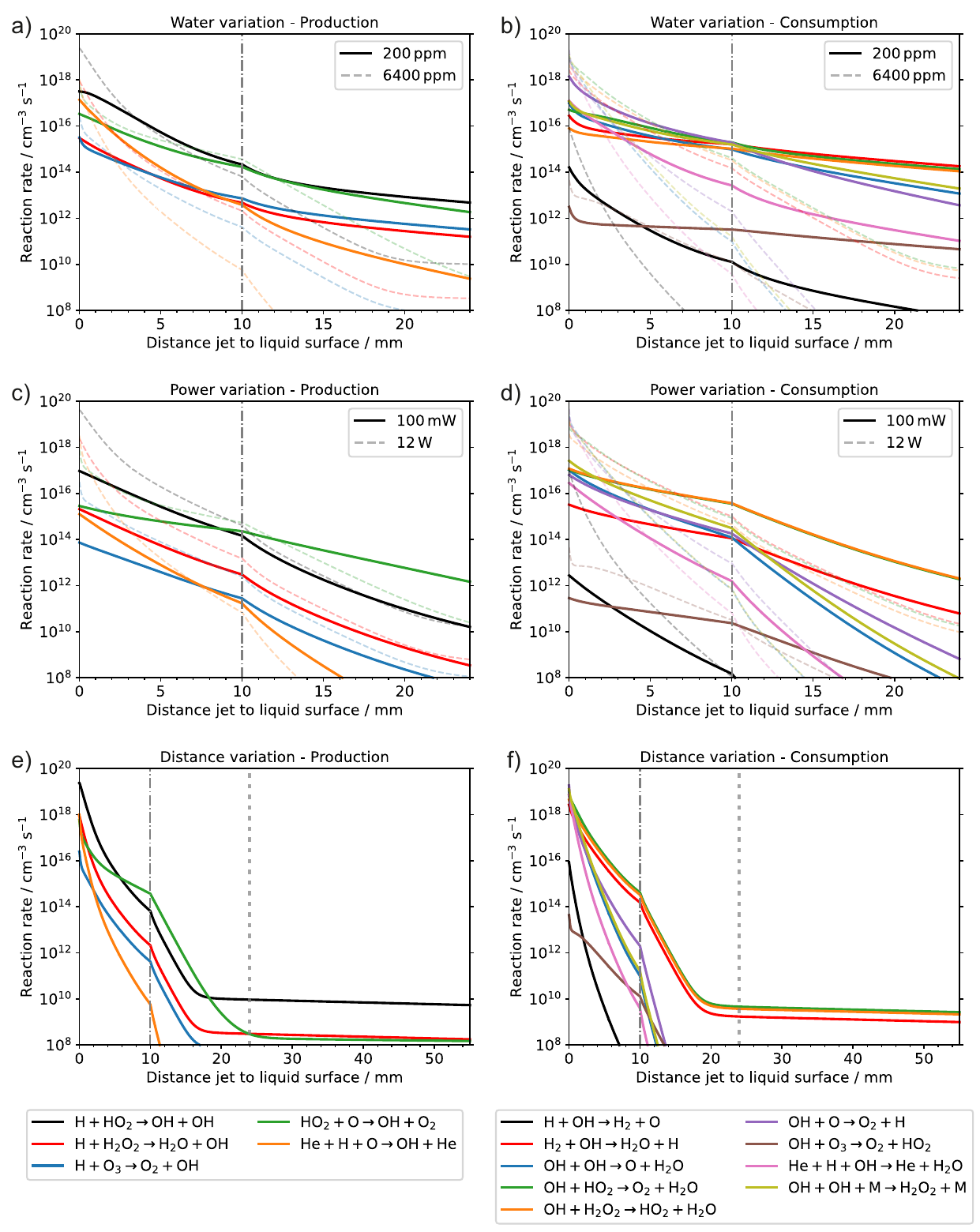}
    \caption{Main production (legend lower left for a), c) and e)) and consumption pathways (legend lower right for b), d) and f)) of OH in the gas phase over distance between the active plasma region and liquid surface under water admixture variation (a) and b)), power variation (c) and d)) and distance variation jet to liquid surface (e) and f)). The vertical dash-dotted line indicates the end of the capillary guided effluent. The vertical dotted line in e) and f) indicates 24\,mm.}
    \label{fig:pathwaysOH}
\end{figure}

Therefore, Fig.~\ref{fig:pathwaysOH} shows the main production and consumption pathways in the effluent region for the water admixture variation (a) and b)), the deposited plasma power variation (c) and d)) and the distance variation (e) and f)). Here, the vertical dash-dotted line at 10\,mm indicates the end of the capillary guided effluent. The solid lines in a) and b) correspond to the lower water admixture where the maximum in the OH density is found, i.e. 200\,ppm. The dashed lines correspond to the saturation water admixture of 6400\,ppm. The solid lines in c) and d) correspond to 100\,mW deposited plasma power, while the transparent dashed lines correspond to 12\,W deposited plasma power. The vertical dotted line in e) and f) indicates the minimal distance of 24\,mm compared to 55\,mm between the end of the capillary plasma jet and the liquid surface.

Overall, it is visible that for all of the parameters shown, consumption dominates over the production of OH since there are more consumption reactions with higher reaction rates than production reactions. This is to be expected in the effluent region, where electrons are not present to drive the dissociation of water vapour.

Even though the density of OH at the exit of the plasma region would be assumed to increase with increasing water admixture, the simulation results in the gas phase at the liquid surface and in the liquid show a decrease above a water admixture of 200\,ppm. The increase of OH with water admixture, is valid for the plasma region, however, it is followed by a decrease of the highly reactive OH in the effluent region \cite{schuttler_production_2024}. Since the OH behaves differently for each parameter, the pathways producing and consuming the OH must change. Most production reactions have a higher rate at the beginning of the effluent region, changing with increasing distance into the effluent. This is because the important production reactions for OH in the effluent generally involve H and O, which themselves are short lived and whose densities are decreasing with increasing distance into the effluent. The main consumption reaction partners of OH in the effluent are typically H$_2$, HO$_2$, H$_2$O$_2$ and OH itself.

For the water admixture variation, the dominant production reaction at 200\,ppm at the start of the effluent is the reaction of H and HO$_2$ producing 2\,OH, followed by O and HO$_2$ producing O$_2$ and OH, and the three body recombination of H and O. In comparison, the main production reaction at 6400\,ppm water admixture varies more significantly over the effluent distance, changing from the reaction of H and HO$_2$ in the near-effluent to that involving O and HO$_2$ at larger distances. Compared to the production of OH, its consumption in the effluent is generally more complex with a larger range of reactions contributing. At 200\,ppm, the main consumption partner for OH at the start of the effluent is O, while further into the effluent reactions with longer lived species such as H$_2$, HO$_2$ and H$_2$O$_2$ begin to become more important. At 6400\,ppm, OH consumption happens at high rates with O, H$_2$, HO$_2$, H$_2$O$_2$ and OH.
Overall, the sum of the consumption reactions for OH at the start of the effluent is significantly higher for the case with higher water admixture (6400\,ppm) due to the presence of higher concentrations of these consumption partners, in comparison to the low water admixture (200\,ppm) case. The result is that the large concentrations of OH present at the end of the plasma region at 6400\,ppm decay more rapidly in the effluent, so that the OH density at the liquid surface is lower than for the 200\,ppm case.

The main OH production pathways under deposited plasma power variation are similar to the water admixture variation. Here, the sum of the production reactions is higher for a deposited power of 12\,W compared to 100\,mW. At the beginning of the effluent region, up to nine to ten millimetres, the reaction between H and HO$_2$ is the dominant one, changing for higher distances to the jet to the reaction between HO$_2$ and O to become the dominant production pathway of OH. Overall, the sum of the main production pathways of OH is higher at 12\,W deposited plasma power than at 100\,mW. In contrast, the consumption of OH is again more complex with a larger number of reactions playing a role. The most important consumption reactions at a plasma power of 100\,mW are OH reacting with HO$_2$ and H$_2$O$_2$, and the three-body recombination between two OH and the feed gas which dominates at the beginning of the effluent. In contrast, there are a larger number of important consumption reactions for 12\,W deposited plasma power, mostly starting two orders of magnitude higher but decaying faster. Here, the two main consumption reactions are OH reacting with H$_2$ and with HO$_2$, followed by the reaction between OH and H$_2$O$_2$, which is one of the two dominant reactions at 100\,mW deposited plasma power. Overall, there is a change in the chemical behaviour under deposited plasma power variation, with the presence of higher densities of consumption reaction partners leading to faster consumption of OH in the effluent at higher powers, and ultimately lower densities of OH at the liquid surface.

The comparison of the OH reaction pathways for the variation of the distance between the capillary plasma jet and the liquid surface is depicted in Fig.~\ref{fig:pathwaysOH}~e) for the production and in Fig.~\ref{fig:pathwaysOH}~f) for the consumption. Here, the pathways over distances of 24\,mm and 55\,mm from the active plasma region are compared. Pathways are shown for the 55\,mm distance simulation only, as the pathways for the 24\,mm simulation behave similarly. A distance of 24\,mm is marked using the vertical dotted line within the graph to visualise the end of the 24\,mm simulation.
As for the cases described above, the production of OH in the effluent mainly depends on reactions between H reacting with HO$_2$ or H$_2$O$_2$, but also of HO$_2$ reacting with O. While the reaction between H and HO$_2$ is the main production pathway at the beginning of the effluent region, HO$_2$ reacting with O becomes dominant above 6\,mm. At 19\,mm, the reaction between H and HO$_2$ becomes dominant again and remains the most important until 55\,mm. The reaction between H and H$_2$O$_2$ becomes more important than that between HO$_2$ and O at 24\,mm. All relevant production reaction rates stay nearly constant at distances above 24\,mm until the liquid surface at 55\,mm. 
The consumption of OH, again, depends on more reactions than the production of OH. Therefore, the OH density decreases while passing the gap between jet and liquid. The main consumption pathways over distance are OH reacting with HO$_2$ followed by H$_2$O$_2$ and H$_2$. These are the only remaining reactions at 24\,mm and stay nearly constant until 55\,mm. At the beginning of the effluent region, the three-body recombination of two OH and the background gas, reactions with H and the background gas, reactions with O, and the two-body reaction of two OH are also relevant. Overall, no significant changes occur in the OH production and consumption dynamics between  24\,mm and 55\,mm, and the density of OH is generally decreasing over this range. These trends are consistent with the decreasing trends of the OH density in the gas phase at the liquid surface, and the concentration within the liquid shown above.

Overall, the pathways show that the consumption of OH in the effluent is dominant in all cases, compared to the production. The main reactants to produce OH are H, O, HO$_2$ and H$_2$O$_2$ with changing importance over parameter variation. The main consumption partners of OH are HO$_2$, H$_2$O$_2$, H$_2$ and OH itself. In general, the complex kinetics of OH consumption in the effluent mean that the OH density in the gas phase at the liquid surface, and ultimately inside the liquid itself is not directly correlated to the density of OH at the end of the plasma region, and the presence of long-lived reactive species in the effluent tends to decrease the amount of OH delivered to the surface of the liquid.

\subsection{Concentration profiles of H$_2$O$_2$ and OH within the liquid} \label{sec:profiles}
In order to understand the results presented in previous sections in more detail, the concentration profiles of H$_{2}$O$_{2}$ and OH in the liquid are analysed in this section.

\subsubsection{Water admixture variation} \label{sec:penDepth_watVar}
\begin{figure}[htb]
    \centering
    \includegraphics[width=1.\textwidth]{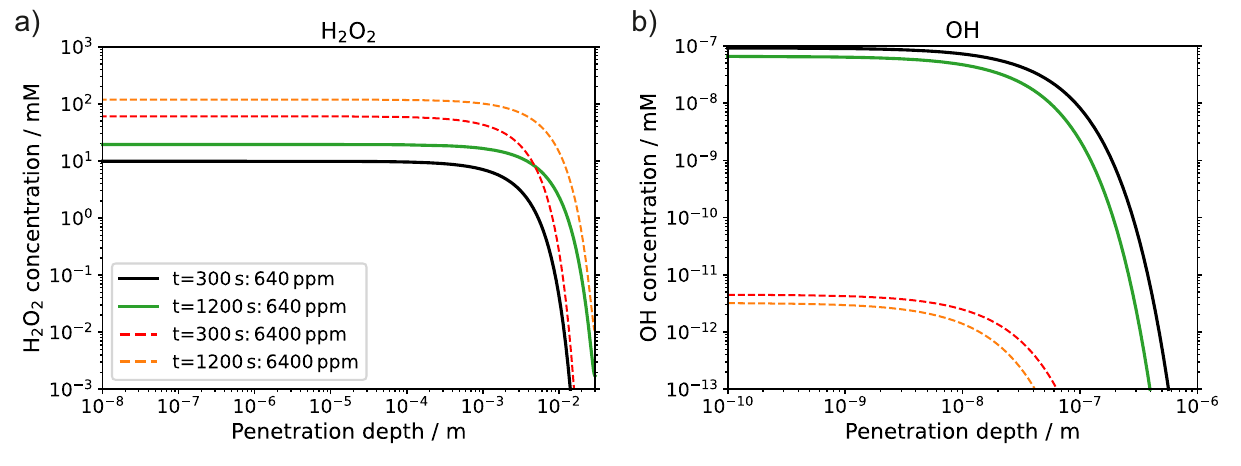}
    \caption{Depth profiles at 300\,s and 1200\,s of plasma treatment of a) H$_2$O$_2$ and b) OH for different water admixtures, a deposited plasma power of 6.0\,W, a total effluent length of 24\,mm, and a total distance from the beginning of the electrodes to the liquid surface of 64\,mm. Legend is true for all.}
    \label{fig:waterVar_depthProfiles}
\end{figure}

The concentration profiles over depth within the liquid are shown in Fig.~\ref{fig:waterVar_depthProfiles} for a) H$_{2}$O$_{2}$ and b) OH, each after 300\,s and 1200\,s plasma treatment time and water admixtures of 640\,ppm and 6400\,ppm. The deposited plasma power and the distance between active plasma region and liquid surface are kept constant at 6.0\,W and 24\,mm.

The H$_{2}$O$_{2}$ concentration is nearly constant up to the millimetre range for both water admixtures after both treatment times, followed by a treatment time-dependent, steep concentration drop. After 1200\,s, the H$_{2}$O$_{2}$ concentration is higher than for 300\,s and penetrates deeper into the liquid. This is caused by the transport of H$_{2}$O$_{2}$, and the presence of a continuous flux from the gas phase. Therefore, longer treatment times increase the concentration and allow more time for transport deeper into the liquid. The treatment time dependence is visible for both water admixtures shown, with higher water admixtures giving higher concentrations of H$_{2}$O$_{2}$. Overall, these results underline the behaviour shown in Fig.~\ref{fig:H2O2_compGas}~a).

The OH concentration is mainly located at the liquid surface, as already shown in Fig.~\ref{fig:contour}~b), and typically penetrates some tens to hundreds of nanometres into the liquid. The OH concentration decreases with increasing treatment time due to the increase in the H$_{2}$O$_{2}$ concentration, which increases the total consumption rate of OH, as discussed earlier. Moreover, compared to the treatment time, the water admixture has a much stronger influence on the OH concentration (four orders of magnitude) in the liquid and also on the penetration depth (one order of magnitude) into the liquid. Both quantities decrease with water admixture as also shown in Fig.~\ref{fig:OH}~a).

Compared to H$_{2}$O$_{2}$, OH has a lower concentration of about eight to thirteen orders of magnitude and penetrates around five orders of magnitude less into the liquid.

\subsubsection{Deposited plasma power variation}
\begin{figure}[htb]
    \centering
    \includegraphics[width=1.\textwidth]{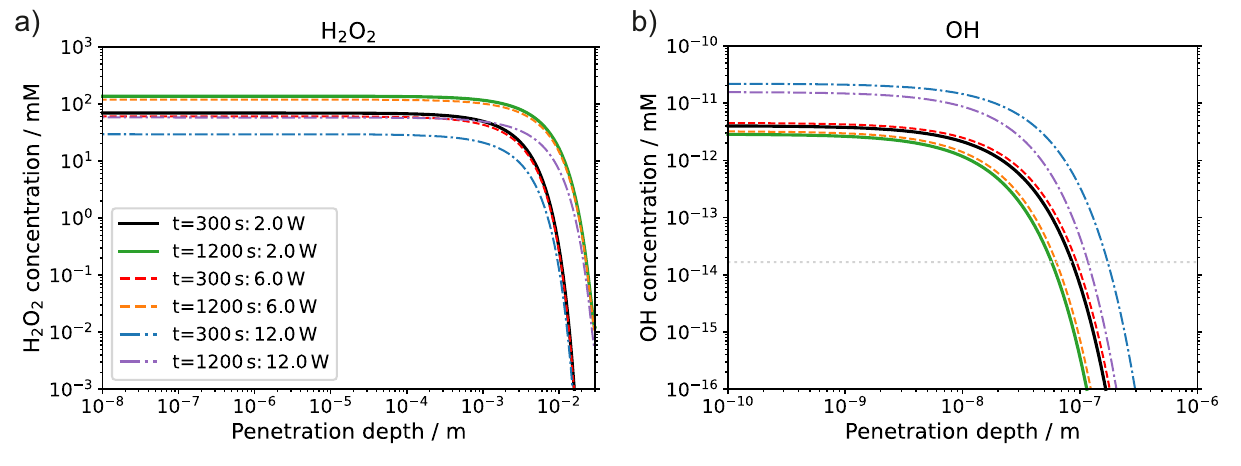}
    \caption{Depth profiles after 300\,s and 1200\,s of plasma treatment of a) H$_2$O$_2$ and b) OH for different deposited power, a water admixture of 6400\,ppm, a total distance from the end of the active plasma region to the liquid surface of 24\,mm, and a total distance from the beginning of the electrodes to the liquid surface of 64\,mm. Legend is true for all.}
    \label{fig:powVar_depthProfiles}
\end{figure}

Fig.~\ref{fig:powVar_depthProfiles} shows the a) H$_{2}$O$_{2}$ concentration and b) OH concentration over depth within the liquid after 300\,s and 1200\,s treatment time at 2.0\,W, 6.0\,W and 12.0\,W deposited plasma power, while water admixture and the distance between the active plasma region and the liquid surface are set constant at 6400\,ppm and 24\,mm.

The H$_{2}$O$_{2}$ behaves similarly in trend over depth within the liquid as described for the water admixture variation in Sec.~\ref{sec:penDepth_watVar} and stays approximately constant up to millimetre range, followed by a steep decrease. The treatment time, again, influences the H$_{2}$O$_{2}$ concentration resulting in an increase from 300\,s to 1200\,s. In the case of the power variation, the H$_{2}$O$_{2}$ concentration slightly decreases from 2.0\,W to 6.0\,W, resulting in a lower penetration depth for both treatment times. The decrease between 6.0\,W and 12.0\,W is more significant, while the penetration depth is comparably less influenced but still decreasing. The power-dependent trend underlines the behaviour shown in Fig.~\ref{fig:H2O2_compGas}~b).

The OH typically penetrates up to hundreds of nanometres into the liquid, with the exact penetration depth depending on the treatment time and the deposited plasma power. While the longer treatment time of 1200\,s causes a decrease of OH compared to a treatment time of 300\,s, higher deposited plasma power causes an increase of the OH concentration. The difference between 2.0\,W and 6.0\,W is less significant than the increase between 6.0\,W and 12\,W. Overall, both treatment time and deposited plasma power cause a noticeable influence on the penetration depth.

Comparing H$_{2}$O$_{2}$ to OH, the H$_{2}$O$_{2}$ has a much higher concentration of around fourteen orders of magnitude and penetrates around five orders of magnitude deeper into the liquid than the OH. Both species vary with treatment time and deposited plasma power but behave contrary to each other over each parameter. The H$_{2}$O$_{2}$ concentration decreases with deposited plasma power, while the OH concentration increases and both species show opposing behaviour with treatment time.

\subsubsection{Variation of the distance between the end of the active plasma region and the liquid surface}
\begin{figure}[htb]
    \centering
    \includegraphics[width=1.\textwidth]{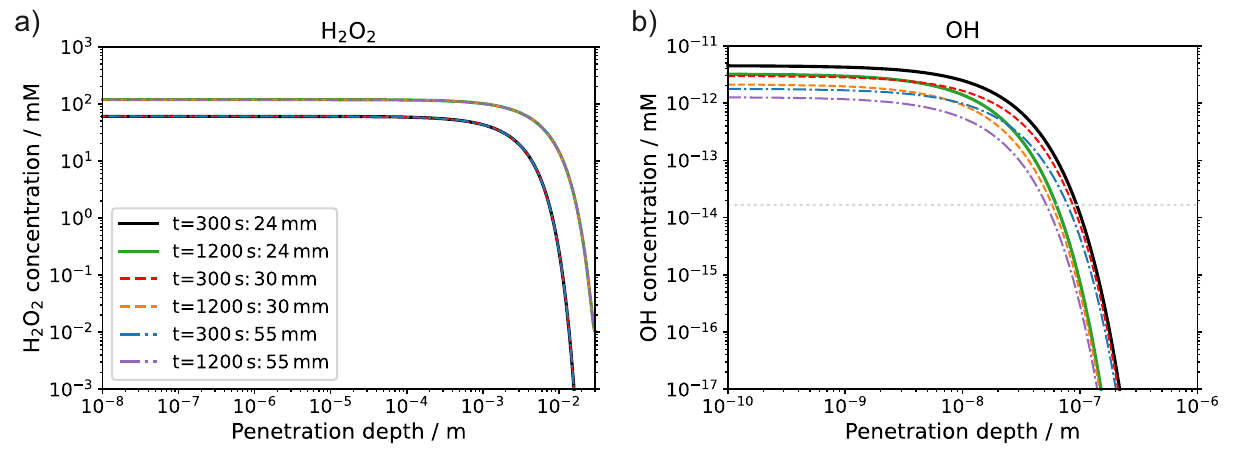}
    \caption{Depth profiles at 300\,s and 1200\,s of plasma treatment of a) H$_2$O$_2$ and b) OH for different distances between the end of the active plasma region and the liquid surface, a deposited plasma power of 6.0\,W, and a water admixture of 6400\,ppm. Legend is true for all.}
    \label{fig:distVar_depthProfiles}
\end{figure}

Here, the influence of the distance between the end of the active plasma region and the liquid surface is investigated. Changing the distance between the active plasma region and the liquid surface only changes the effluent region because the geometry of the jet configuration, including the capillary, is kept constant. Therefore, only the unguided effluent length is varied between 14\,mm and 45\,mm, which is equal to a variation in the distance between the active plasma region and the liquid surface of 24\,mm to 55\,mm and is shown in Fig.~\ref{fig:distVar_depthProfiles}~a) for H$_{2}$O$_{2}$ and b) for OH.

Again, the H$_{2}$O$_{2}$ concentration is approximately constant between the liquid surface and depths up to the millimetre range, followed by a steep decrease. The H$_{2}$O$_{2}$ concentration in the liquid increases with treatment time. In contrast, the variation of the distance between the capillary plasma jet and the liquid surface does not influence the H$_{2}$O$_{2}$ concentration in the model and underlines the behaviour shown in Fig.~\ref{fig:H2O2_compGas}~c). Since H$_{2}$O$_{2}$ is a long-living and stable species, a longer effluent region in this range does not affect it. The chemistry taking place in the liquid does also not affect the H$_{2}$O$_{2}$ in a significant way since the amount of OH is comparably low and therefore, the H$_{2}$O$_{2}$ concentration profile is mainly transport-driven. It is worth to emphasise that the density trends for H$_{2}$O$_{2}$ differ between simulation and experiment, and these simulation results should be interpreted in that context.

The OH again penetrates up to hundreds of nanometres into the liquid and is influenced by the treatment time and the distance between the capillary plasma jet and the liquid surface. Increasing treatment time leads to a decrease of the OH concentration, which is similar to the behaviour with increasing distance between the capillary plasma jet and the liquid surface.

Therefore, comparing H$_{2}$O$_{2}$ and OH, the H$_{2}$O$_{2}$ concentration is, again, around fourteen orders of magnitude higher than of the OH, while H$_{2}$O$_{2}$ penetrates around five orders of magnitude deeper into the liquid up to centimetre range. The H$_{2}$O$_{2}$ is not recognisably influenced by the distance between the capillary plasma jet and the liquid surface, while OH decreases with the distance (solid lines $\rightarrow$ dashed lines $\rightarrow$ dash-dot lines). The behaviour of H$_{2}$O$_{2}$ and OH underlines the behaviour shown in Fig.~\ref{fig:H2O2_compGas}~c) and Fig.~\ref{fig:OH}~c). A longer treatment time of 1200\,s results in a larger concentration of H$_{2}$O$_{2}$ and a lower concentration of OH than a treatment time of 300\,s.

\section{Conclusion} \label{sec:conclusions}
The behaviour of H$_{2}$O$_{2}$ and OH in a liquid environment treated by a capillary plasma jet, operated using a gas flow of 0.25\,slm, has been investigated using a combination of gas and liquid phase models. It is found that the concentration profiles of each species can be tailored by choosing the operating conditions of the jet system. H$_{2}$O$_{2}$ in the liquid behaves as a long-living species, the concentration profiles of which are mainly transport-driven. In contrast, the highly reactive OH behaves as transport-driven until the millisecond range, after which consumption during reactions with H$_{2}$O$_{2}$ limit its concentration and penetration depth. The water admixture, the deposited plasma power and the distance between the jet and the liquid surface are possible parameters influencing the species' behaviour. The simulated trends of the H$_{2}$O$_{2}$ concentration in the liquid are in good agreement with the trends of the experimental measurements for variation of plasma power and water admixture. However, the modelling approach used here does not replicate the experimentally measured trend in H$_{2}$O$_{2}$ concentration with distance between the jet and the liquid surface. The simulation overestimates the experimentally measured absolute values by a factor of five. The difference in experiment and simulation is likely caused by simplifying assumptions in the model, where effects such as the widening and structure of the gas flow on to the liquid surface, mixing with ambient air and transport to the sides within the liquid are neglected. The H$_{2}$O$_{2}$ and OH concentrations in the liquid largely follow their trends in the gas phase. The H$_{2}$O$_{2}$ is strongly influenced by the water admixture and the deposited plasma power, and in the experiment by the distance from the jet to the liquid surface, although the latter trend is not properly captured by the model. The OH is mostly influenced by the variation of the water admixture, followed by the deposited plasma power and the distance between the jet and the liquid. Analysis of reaction pathways in the effluent of the jet showed that OH is consumed within the effluent region since there are fewer production reactions with comparable or lower production reaction rates than consumption reactions. As a result, the OH concentration reaching the surface strongly depends on the densities of various species in the effluent, which can be tailored by the plasma operating conditions. Importantly, this means that it cannot be assumed that the trends in OH densities at the end of the plasma forming region are the same as those at the liquid surface, or inside the liquid.

Moreover, the concentrations and penetration depths of H$_2$O$_2$ and OH were found to differ with increasing plasma treatment time. Increasing treatment time leads to a build up of H$_{2}$O$_{2}$ in the liquid, and a continuous increase in its penetration depth due to transport. On the other hand, the concentration and penetration depth of OH tends to decrease with increasing treatment time, as it is consumed during reactions with H$_{2}$O$_{2}$, whose density is increasing.

Taken together, the results also demonstrate that the relative concentrations of OH and H$_{2}$O$_{2}$ in the liquid can be controlled over orders of magnitude by varying the plasma operating conditions. The highest relative contribution of OH is found at low water admixtures, low powers, and short treatment times, while H$_{2}$O$_{2}$ is favoured at high water admixtures, high powers, and long treatment times.

It should be noted here that the 0.25\,slm gas flow rate used in this work is lower than that typically used in the capillary jet, and other similar jets, such as the COST jet, where a value of 1\,slm is typical. At higher gas flows, the rate of H$_2$O$_2$ transport in the liquid will be higher due to increased gas flow induced liquid convection, and the residence time of species in the effluent will be shorter. While the qualitative trends demonstrated in this work are still expected to be valid for jets operated at higher gas flows, the quantitative details will differ. Specifically, H$_2$O$_2$ would be expected to reach the same penetration depths more quickly at higher gas flow rates, and the relative importance of effluent chemistry for the consumption of OH between the active plasma region and the liquid surface would be expected to decrease.

Overall, the results presented here give insights into how the concentrations of H$_{2}$O$_{2}$ and OH in plasma treated liquids may be tailored by the plasma operating conditions and the plasma treatment time. In this context, these results are relevant for applications where these species are important, such as plasma-driven biocatalysis, chemical conversion and plasma medicine.

For plasma-driven biocatalysis, where a low concentration of OH and a high concentration of H$_2$O$_2$ is favourable, a longer treatment time should be chosen to increase the H$_2$O$_2$ concentration while simultaneously decreasing the amount of OH. Furthermore, lower powers and higher water admixtures should allow for more effective biocatalytic processes. Also, a higher distance between the capillary plasma jet and the liquid surface may be preferable if the OH concentration still needs to be decreased since the OH concentration is influenced more strongly by the distance variation than the H$_2$O$_2$ even in the experimental measurements.

\section{Acknowledgment}
The authors thank Prof. Mark Kushner for providing the GlobalKin code and the German Research Foundation (DFG) for financial support via CRC 1316 (project number 327886311), project B11.


\newcommand{\newblock}{}

\end{document}